\documentclass{article}

\usepackage{mathptmx}
\usepackage{cite}
\usepackage{graphicx}
\usepackage{xcolor}
\usepackage{hyperref}
\usepackage{amsmath, amsthm, amssymb, amsfonts,amsbsy,amstext,amscd,amsxtra,amsopn,dsfont}
\usepackage[utf8]{inputenc}
\usepackage{braket}
\usepackage{float}
\usepackage[english]{babel}
\usepackage{subfig}
\usepackage[top=2cm,bottom=2cm,right=2cm,left=2cm]{geometry}

\begin{document} 

\title{Effect of the nearby levels on the resonance fluorescence spectrum of the atom-field interaction}

\author{L. Villanueva-Vergara, F. Soto-Eguibar,H. M. Moya-Cessa\\
\small{Instituto Nacional de Astrofísica, Óptica y Electrónica, INAOE}\\
\small{Luis Enrique Erro 1, Santa María Tonantzintla, San Andrés Cholula, Puebla, México 72840}}

\maketitle

\begin{abstract}
We study the resonance fluorescence in the Jaynes-Cummings model when nearby levels are taking into account. We show that the Stark shift produced by such levels generates a displacement of the peaks  of the resonance fluorescence due to an induced effective detuning and also induces an asymmetry. Specific results are presented assuming a coherent and a thermal fields.
\end{abstract}

\section{Introduction.}
The Jaynes-Cummings model has been extensively studied through the years; it considers the interaction of a two-level atom and a single field cavity mode in the dipole approximation \cite{JaynesCummings1963,knight,moyalibro,Swain1972}. This model not only can be exactly solved, but it has interesting features such as collapse and revivals of Rabi oscillations and spontaneous emission \cite{EberlyNarozhny1980,NarozhnyEberly1981}. On the other hand, the AC Stark shift or dynamical Stark effect occurs when an optical field interacts with an atom, and in consequence, their energy levels are shifted. In Ref. \cite{Meschede1985}, it was shown that the spectral lines of the Rubidium isotope $^{85}$Rb are asymmetric and shifted; this behavior was attributed to the presence of non-resonant energy levels. In Ref.\cite{Moya1991}, it was shown that the effect of nearby levels could be effectively seen as an AC Stark shift in the dynamics of the excited, $ \ket{e}, $ and  ground, $ \ket{g}$, states; to phenomenologically introduce that interaction, the term $\chi\hat{n}\sigma_z$ was added to the Jaynes-Cummings Hamiltonian. The phenomenological Hamiltonian that describes the atom-field interaction taking into account the rotating wave approximation (RWA) and the presence of higher nearby levels may be written as  
\begin{equation}\label{0010}
\hat{H}_p=\omega\hat{n}+\frac{\omega_0}{2}\hat{\sigma}_z+\lambda\left(\hat{a}\hat{\sigma}_++\hat{\sigma}_-\hat{a}^\dagger \right)+\chi\hat{n}\hat{\sigma}_z.
\end{equation}
where the operators $\hat{\sigma}_{\pm,z} $ are the atomic operators and $ \hat{a}$, $ \hat{a}^\dagger $ and $ \hat{n} $ are field operators defined as usual (see the Hamiltonian below for the definition of all the parameters). The parameter $\chi$ represents the off-resonant interaction between the quantized field and the nearby levels.\\
The fluorescence spectrum of a two-level atom interacting with a single-mode field of a cavity has been extensively analyzed \cite{Mondragon1983,KnightMilonni1980}. One way to solve the dynamics of the system is through Bloch optical equations \cite{AllenEberly1975}; however, in \cite{Agarwal1985} a method was used which expresses the evolution operator in terms of the dressed states that, in addition, have already been shown to diagonalize the Hamiltonian of the Jaynes-Cummings model \cite{Tannoudji1975}. In Ref. \cite{Eberly1977} the physical spectrum for a non stationary processes was found; this method has been used also to find the fluorescence spectrum of the Jaynes-Cummings model \cite{Agarwal1991} and of a model that includes a Kerr medium in a cavity \cite{JoshiLawande1992}. \\
In this contribution, we  start from a Hamiltonian that describes the interaction between an atom with a ground state $\ket{g}$, an excited state $ \ket{e} $ and  N states $\ket{k} $ with $ k = 1, .. ., N $ and one-mode quantized field. The coupling between $\ket{g}$ and $ \ket{k} $ is weak. Employing small rotation transformations \cite{Klimov2000}, we formally cast  the general Hamiltonian into an effective Hamiltonian that has an AC Stark term. Although the phenomenological model \eqref{0010} incorporates Stark shifts in a proper way, this expression had not been formally demonstrated from a model that includes the interaction between the field  and the non-resonant but nearby states $\ket{k}$; here we achieve that goal.\\
Once established the validity of the Hamiltonian, eigenstates and eigenenergies are calculated under the dressed atom picture and the unitary evolution operator $\hat{U}_\mathrm{SE}(t)$ is found. We calculate the correlation function $ \Gamma(t,\tau)=\braket {\hat {\sigma}_+(t+\tau)\hat{\sigma}_-(t)} $ and its Fourier transform, so we obtain an expression for the emission spectrum. Concrete results of the fluorescence spectrum are shown, considering different initial conditions for the field, such as coherent and thermal states.\\
	
\section{The Jaynes-Cummings model plus nearby levels}
\begin{figure}[H]
\centering
\includegraphics[width=0.5\textwidth]{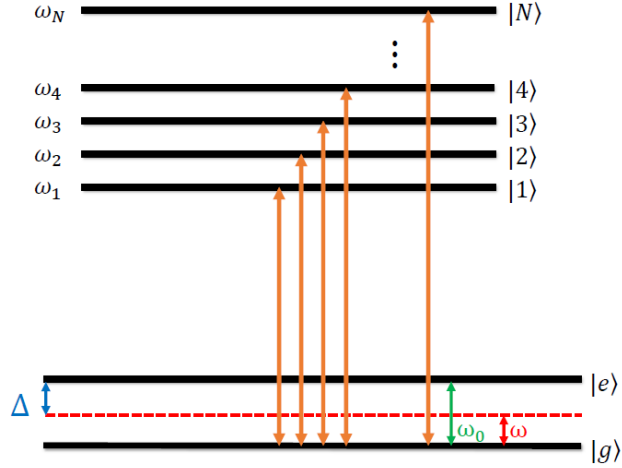}
\caption{Level scheme of atomic states. The transition frequency between $\ket{g}$ and $\ket{e}$ is denoted by $\omega_0$, the field frequency is $\omega$ and $\omega_k$ are the transition frequencies between off-resonant states $\ket{k}$ and $\ket{g}$ with $k=1,...,N$.}\label{fig1}
\end{figure}

Consider an atom with a ground state $\ket{g}$, an excited state $\ket{e}$ and $N$ higher states denoted by $\ket{k}$, where $k=1,2,3,...N$. The atom is interacting with a single mode field, as shown in Fig. \ref{fig1}. The quantized field is slightly detuned from the two lower levels of the atom and highly detuned from nearby levels $\ket{k}$. The Hamiltonian representing the complete system is given by
\begin{equation}\label{0020}
	\hat{H}=\omega\hat{n}+\frac{\omega_0}{2}\hat{\sigma}_z+\frac{1}{2}\sum_{k=1}^N\omega_k\ket{k}\bra{k}
	+\lambda\left(\hat{a}\hat{\sigma}_+	+\hat{a}^\dagger\hat{\sigma}_- \right)+\sum_{k=1}^N\eta_k\left(\hat{a}\ket{k}\bra{g}+\hat{a}^\dagger\ket{g}\bra{k} \right),
\end{equation}
where $\omega$ is the frequency of the single mode quantized field, $\omega_0$ is the atomic transition frequency between states $\ket{g}$ and $\ket{e}$, and $\omega_k$ are the frequencies of the high nearby states $\ket{k}, \; k=1,2,3,...N$. The field operators involved in the Hamiltonian are the field photon annihilation $\hat{a}$ and creation $\hat{a}^\dagger$ operators, and the field photon number $\hat{n}=\hat{a}^\dagger \hat{a}$ operator. For the atom, we have the Pauli $z$ matrix $\hat{\sigma}_z=\ket{e}\bra{e}-\ket{g}\bra{g}$ and the raising $\hat{\sigma}_+=\ket{e}\bra{g}$ and lowering $\hat{\sigma}_-=\ket{g}\bra{e}$ operators between the lower states, $\ket{g}$ and $\ket{e}$. The coupling between the field and the atom is measured by the coupling constants: $\lambda$ gives the strength of the coupling between the field and the atom when it is in the two lower states, and $\eta_k, \; k=1,2,3,...N$ are the coupling constants of the transitions between the $\ket{g}$ state and the $\ket{k}$ states. We will consider that these last interactions are much smaller than the two-level transition $\ket{g}\leftrightarrow \ket{e}$. In addition, we will also suppose that $\omega_k \gg \omega_0$ for all $k$ from $1$ to $N$.
\noindent In order to simplify the Hamiltonian \eqref{0020}, we transform it with the unitary rotation
\begin{equation}
\hat{R}=\exp\left[ \sum_{j=1}^{N}  \xi_j(\hat{a}\ket{j}\bra{g}-\hat{a}^\dagger\ket{g}\bra{j})\right] ,
\end{equation}
with $\xi_j\ll 1$ for $j=1,2,3,...,N$; as all the parameters involved in the rotation are small, we will call it an small rotation. These parameters will be fixed later (Eq. \eqref{0050}) and will allow us to neglect terms that exchange energy between the quantized field and the high nearby levels. Neglecting terms of order $\xi^2$ or higher, we obtain, after a straightforward but long calculation, that
\begin{align}
\hat{H}_{\mathrm{rot}}&=\hat{R}\hat{H}\hat{R}^\dagger\simeq  \omega\hat{n}+\frac{1}{2}\omega_0\hat{\sigma}_z+\lambda\left(\hat{a}\hat{\sigma}_++\hat{a}^\dagger\hat{\sigma}_- \right)+\frac{1}{2}\sum_{k=1}^N\omega_k\ket{k}\bra{k}
\nonumber \\ &
+\lambda\left(\hat{n}+1 \right)\sum_{k=1}^N\xi_k\left(\ket{e}\bra{k}+\ket{k}\bra{e} \right)+\left(\hat{n}+1 \right)\sum_{j,k=1}^N\xi_j\eta_k\left(\ket{k}\bra{j}+\ket{j}\bra{k} \right)
\nonumber \\ &
-2\hat{n}\sum_{k=1}^N\xi_k \eta_k \ket{g}\bra{g}
+\sum_{k=1}^N\left[\xi_k\left(\omega-\frac{\omega_k+\omega_0}{2} \right)+\eta_k \right]\left( \hat{a}\ket{k}\bra{g}+\hat{a}^\dagger\ket{g}\bra{k}\right).
\end{align}
To cancel the last term in the previous Hamiltonian, we fix the parameters $\xi_j, \, j=1,2,3,...,N$ as
\begin{equation}\label{0050}
\xi_j=\frac{2\eta_j}{\Delta+\Delta_j}, \qquad j=1,2,3,\dots,N,
\end{equation}
where the detunings are $\Delta=\omega_0-\omega$ and $\Delta_j=\omega_j-\omega$. Note that the conditions $\xi_j\ll 1$ for $j=1,2,3,...,N$ requires that  the coupling constants and the frequencies satisfy the relations
\begin{equation}\label{0060}
\eta_j \ll	\frac{\Delta_j+\Delta}{2}, \qquad j=1,2,3,\dots,N.
\end{equation}
With this choice of the parameters of the small rotation, we arrive to
\begin{align}\label{0070}
\hat{H}_{\mathrm{rot}}=&\Omega\hat{n}+\frac{\omega_0}{2}\hat{\sigma}_z+\frac{1}{2}\sum_{k=1}^N\omega_k\ket{k}\bra{k}+\lambda\left(\hat{a}\hat{\sigma}_+ +\hat{a}^\dagger\hat{\sigma}_-\right)+\chi\hat{n}\hat{\sigma}_z
+\chi\hat{n} \sum_{k=1}^N\ket{k}\bra{k} 
\nonumber\\ &
+\lambda\left(\hat{n}+1 \right)\sum_{k=1}^N\xi_k\left(\ket{e}\bra{k}+\ket{k}\bra{e} \right)
+\left(\hat{n}+1 \right)\sum_{j,k=1}^N\xi_j\eta_k\left(\ket{k}\bra{j}+\ket{j}\bra{k} \right),
\end{align} 
where we have defined
\begin{equation}\label{0080}
\chi \equiv \sum_{j=1}^N\xi_j\eta_j, \qquad		\Omega \equiv \omega-\chi.
\end{equation}
We want to find the effective effect of the high nearby states in the interaction; as we show next, this effect is similar to the Stark effect. To achieve this, we transform the rotated Hamiltonian \eqref{0070} to a partial interaction picture \cite{roman} by means of the unitary transformation
\begin{equation}
\hat{T}=\exp\left[ -it\left(\frac{1}{2}\omega_0\hat{\sigma}_z+\frac{1}{2}\sum_{k=1}^N\omega_{k}\ket{k}\bra{k}\right)\right],
\end{equation}
obtaining for the transformed Hamiltonian of the corresponding transformed Schrödinger equation
\begin{align}
\hat{H}_\mathrm{I}=
&\Omega\hat{n}+\lambda\left[\hat{a}\hat{\sigma}_+ \exp\left( i \omega_0 t \right)  +\hat{a}^\dagger\hat{\sigma}_- \exp\left( -i \omega_0 t \right) \right] +\chi\hat{n}\hat{\sigma}_z
+\chi\hat{n}\sum_{k=1}^N\ket{k}\bra{k}
\nonumber\\	&
+\left(\hat{n}+1\right)\sum_{j,k=1}^N\xi_k\eta_j
\left[\exp\left(i t \frac{\omega_j - \omega_k}{2}\right)\ket{j}\bra{k}+ \exp\left(-i t \frac{\omega_j - \omega_k}{2} \right) \ket{k}\bra{j} \right]
\nonumber\\	&
+\lambda\left(\hat{n}+1\right)\sum_{k=1}^N\xi_k\left[\exp\left(-i t \frac{\omega_k - \omega_0}{2}\right)\ket{e}\bra{k}+ \exp\left(i t \frac{\omega_k - \omega_0}{2} \right) \ket{k}\bra{e} \right].
\end{align}
As we have supposed that $\omega_k\gg\omega_0, k=1,2,3 \dots, N$, the terms $\exp\left(\pm i t \frac{\omega_k - \omega_0}{2} \right)$ oscillate very fast and we can make a second rotating wave approximation despising those terms and obtaining
\begin{align}
&\hat{H}_\mathrm{RWA}=
\Omega\hat{n}+\lambda\left[\hat{a}\hat{\sigma}_+ \exp\left( i \omega_0 t \right)  +\hat{a}^\dagger\hat{\sigma}_- \exp\left( -i \omega_0 t \right) \right] +\chi\hat{n}\hat{\sigma}_z
+\chi\hat{n}\sum_{k=1}^N\ket{k}\bra{k}
\nonumber\\	&
+\left(\hat{n}+1\right)\sum_{j,k=1}^N\xi_k\eta_j
\left[\exp\left(i t \frac{\omega_j - \omega_k}{2}\right)\ket{j}\bra{k}+ \exp\left(-i t \frac{\omega_j - \omega_k}{2} \right) \ket{k}\bra{j} \right].
\end{align}
We need to get rid out of time in the second term of this last Hamiltonian; in order to do this, we go back to a Schrödinger type picture by means of the unitary transformation
\begin{equation}
\hat{S}=\exp\left(i t \frac{\omega_0}{2} \hat{\sigma}_z \right) . 
\end{equation}
The Hamiltonian that corresponds to the transformed Schrödinger equation is
\begin{equation}
\hat{H}_\textrm{eff}=\hat{H}_{SE}+\hat{\mathcal{H}},
\end{equation}
where
\begin{equation}\label{0140}
\hat{H}_\mathrm{SE}=\Omega \hat{n}  +\frac{\omega_0}{2} \hat{\sigma}_z
+\lambda\left( \hat{a}\hat{\sigma}_+  + \hat{a}^\dagger\hat{\sigma}_-\right) 
+ \chi \hat{n} \hat{\sigma}_z,
\end{equation}
denoting $\mathrm{SE}$ Stark effect, and being
\begin{equation}
\hat{\mathcal{H}}=
\chi\hat{n}\sum_{k=1}^N\ket{k}\bra{k}
+\left(\hat{n}+1\right)\sum_{j,k=1}^N\xi_k\eta_j
\left[\exp\left(i t \frac{\omega_j - \omega_k}{2}\right)\ket{j}\bra{k}+ \exp\left(-i t \frac{\omega_j - \omega_k}{2} \right) \ket{k}\bra{j} \right].
\end{equation}
It is straightforward to prove that $[\hat{H}_\mathrm{SE},\hat{\mathcal{H}}]=0$, therefore, exists a basis that simultaneously diagonalize $\hat{H}_\mathrm{SE}$ and $\hat{\mathcal{H}}$ and they act in independent subspaces. This result allows $\hat{H}_\mathrm{SE}$ to be considered as effective interaction Hamiltonian since it contains the parameter that represents the effect of nearby levels.
	
\section{The fluorescence spectrum}
We will now calculate the fluorescence spectrum for the system represented by the Hamiltonian $\hat{H}_\mathrm{SE}$, Eq. \eqref{0140}, using the expression for the physical spectrum of a non-stationary process developed by Eberly and Wodkiewicz in \cite{Eberly1977}
\begin{equation}
	S(\nu)=\mathrm{Re}\left\{\int_{0}^\infty e^{-i\nu\tau}e^{-\gamma\tau}\bar{\Gamma}(\tau)d\tau\right \},
\end{equation}
where $\gamma$ represents the width of the detector and $\bar{\Gamma}(\tau)$ is the one cycle time average of the two time correlation function
\begin{equation}
\Gamma(t,\tau)=\braket{\hat{\sigma}_+(t+\tau)\hat{\sigma}_-(t)}.
\end{equation}
To achieve this goal, we need the evolution operator of the system, which we will find using the dressed states. First, we rewrite the Hamiltonian \eqref{0140} as a free part, $\hat{H}_f$, and an interaction part, $\hat{H}_i$, as
\begin{equation}
\hat{H}_\mathrm{SE}=\hat{H}_f+\hat{H}_i,
\end{equation}
with
\begin{equation}
\hat{H}_f=\Omega \hat{n}  +\frac{\omega_0}{2} \hat{\sigma}_z
+ \chi \hat{n} \hat{\sigma}_z
\end{equation}
and
\begin{equation}
\hat{H}_i=\lambda\left( \hat{a}\hat{\sigma}_+  +\hat{a}^\dagger\hat{\sigma}_-\right).
\end{equation}
The bare states $\ket{n,g}$ and $\ket{n,e}$ are eigenstates of $\hat{H}_f$, but $\hat{H}_i \ket{n,e}=\lambda \sqrt{n+1}\ket{n+1,g}$ and $\hat{H}_i \ket{n+1,g}=\lambda \sqrt{n+1}\ket{n,e}$; thus, the dynamics is restricted to the subspace generated by $\ket{n,e}$ and $\ket{n+1,g}$ and the Hilbert space is composed by orthogonal subspaces. Applying then the standard procedure \cite{knight} the dressed states are calculated, finding
\begin{subequations}
\begin{align}
	\ket{\psi_n^+}&=\cos\Phi_n\ket{n,e}+\sin\Phi_n\ket{n+1,g},
	\\
	\ket{\psi_n^-}&=-\sin\Phi_n\ket{n,e}+\cos\Phi_n\ket{n+1,g},
\end{align}
\end{subequations}
with eigenenergies given by
\begin{equation}
E_n^\pm=\omega\left(n+\frac{1}{2}\right)-\frac{\chi}{2}\pm\frac{\mu_n}{2}.
\end{equation}
and where the following quantities have been defined 
\begin{subequations}
\begin{align}
	\Phi_n&=\arctan \left(\frac{\Omega_n}{\mu_n+\delta_n}\right),\\
	\Omega_n&=2\lambda\sqrt{n+1},\\
	\delta_n&=\Delta+\chi(2n+1),\\
	\mu_n&=\sqrt{\left[\Delta+\chi(2n+1)\right]^2+4\lambda^2(n+1)}=\sqrt{\delta_n^2+\Omega_n^2}.
\end{align}
\end{subequations}
It is important to note that the ground state $ \ket{0, g}$ is also an eigenstate of $ \hat{H}_\mathrm{SE}$ with eigenvalue $-\omega_0/2$. The closure relation reads as
\begin{equation}
\ket{0,g}\bra{0,g}+\sum_{n=0}^\infty\left( \ket{\psi_n^+}\bra{\psi_n^+} + \ket{\psi_n^-}\bra{\psi_n^-}\right)=\hat{I} .
\end{equation}
and the evolution operator $\hat{U}_\mathrm{SE}=\exp\left( -i t \hat{H}_\mathrm{SE}\right) $ can be cast, using the closure relation, as
\begin{align}
\hat{U}_\mathrm{SE}=&\exp\left( -i t \hat{H}_\mathrm{SE}\right)
=\exp\left( -i t \hat{H}_\mathrm{SE}\right)\hat{I}
\nonumber \\ 
=&\exp\left( -i t \hat{H}_\mathrm{SE}\right)
\left[ \ket{0,g}\bra{0,g}+\sum_{n=0}^\infty\left( \ket{\psi_n^+}\bra{\psi_n^+} + \ket{\psi_n^-}\bra{\psi_n^-}\right)\right] 
\nonumber \\ 
=&e^{it\omega_0/2}\ket{0,g}\bra{0,g}
+\sum_{n=0}^\infty\left[ D_n(t)\ket{n,e}\bra{n,e}+F_n(t)\left(\ket{n,e}\bra{n+1,g}+\ket{n+1,g}\ket{n,e}\right)
+G_n(t)\ket{n+1,g}\bra{n+1,g}\right],
\end{align}
where the functions introduced above are given by
\begin{subequations}
\begin{align}
		D_n(t)=\exp\left( -i t E_n^+\right) \cos^2\Phi_n+\exp\left( -i t E_n^-\right) \sin^2\Phi_n,\label{30a}\\
		F_n(t)=\cos\Phi_n\sin\Phi_n\left[ \exp\left( -i tE_n^+\right)-\exp\left( -i tE_n^-\right)\right] ,\label{30b}\\
		G_n(t)=\exp\left( -i tE_n^+\right)\sin^2\Phi_n+\exp\left( -i tE_n^-\right)\cos^2\Phi_n.\label{30c}
\end{align}
\end{subequations}
We suppose now that initially the atom is in the excited state $\ket{e}$ and the field in an arbitrary state characterized by the probabilities $p_{m,m'}$, in such a way that the initial density matrix is
\begin{equation}
\hat{\rho}_0=\sum_{m,m'=0}^{\infty} p_{m,m'} \ket{m,e}\bra{m',e};
\end{equation}
so,
\begin{align}\label{33}
\Gamma(t,\tau)=&\braket{\hat{\sigma}_+(t+\tau)\hat{\sigma}_-(t)}
=\mathrm{Tr}\{\hat{\rho}_0\hat{\sigma}_+(t+\tau)\hat{\sigma}_-(t) \}\nonumber\\
=&\sum_{m=0}^\infty\sum_{m'=0}^\infty p_{m,m'}\braket{m',e|\hat{\sigma}_+(t+\tau)\hat{\sigma}_-(t)|m,e}\nonumber\\
=&\sum_{m=0}^\infty\sum_{m'=0}^\infty p_{m,m'}\braket{m',e|e^{i(t+\tau)\hat{H}_{\mathrm{SE}}}\hat{\sigma}_+e^{-i\tau\hat{H}_{\mathrm{SE}}}\hat{\sigma}_-e^{-it\hat{H}_{\mathrm{SE}}}|m,e}.
\end{align}
After making the necessary calculations, we find
\begin{align}
\Gamma(t,\tau)=&p_{0,0}\big\{e^{i\tau(E_0^++\omega_0/2)}\cos^4\Phi_0+e^{i\tau(E_0^-+\omega_0/2)}\sin^4\Phi_0
\nonumber\\&
+\cos^2\Phi_0\sin^2\Phi_0[e^{i\tau(E_0^-+\omega_0/2)}e^{-it\mu_0}+e^{i\tau(E_0^++\omega_0/2)}e^{it\mu_0}] \big\}
\nonumber\\&
+\sum_{m=1}p_{m,m} \bigg\{ e^{i\tau(E_m^+-E_{m-1}^+)}\cos^4\Phi_m\sin^2\Phi_{m-1}+e^{i\tau(E_m^+-E_{m-1}^-)}\cos^4\Phi_m\cos^2\Phi_{m-1}
\nonumber\\&
+e^{i\tau(E_m^--E_{m-1}^+)}\sin^4\Phi_m\sin^2\Phi_{m-1}+e^{i\tau(E_m^--E_{m-1}^-)}\sin^4\Phi_m\cos^2\Phi_{m-1} 
\nonumber\\&
+\cos^2\Phi_m\sin^2\Phi_m\sin^2\Phi_m\left[e^{it\mu_m}e^{i\tau(E_m^+-E_{m-1}^+)}+e^{-it\mu_m}e^{i\tau(E_m^--E_{m-1}^+)} \right]
\nonumber\\&
+\cos^2\Phi_m\sin^2\Phi_m\cos^2\Phi_m\left[e^{it\mu_m}e^{i\tau(E_m^+-E_{m-1}^-)}+e^{-it\mu_m}e^{i\tau(E_m^--E_{m-1}^-)} \right]\bigg\}.
\end{align}
It is clear that this function depends on $t$ and $\tau$, which indicates that it is not a stationary process. Performing an average over one period $T$, $\bar{\Gamma}\left(\tau \right)=\frac{1}{T} \int_{0}^{T} \Gamma\left(t,\tau\right) dt$, we get
\begin{align}
\bar{\Gamma}(\tau)=&p_0^2\left[e^{i\tau(E_0^++\omega_0/2)}\cos^4\Phi_0+e^{i\tau(E_0^-+\omega_0/2)}\sin^4\Phi_0 \right]
\nonumber\\&
+\sum_{m=1}^\infty p_m^2 \big[ e^{i\tau(E_m^+-E_{m-1}^+)}\cos^4\Phi_m\sin^2\Phi_{m-1}+e^{i\tau(E_m^--E_{m-1}^+)}\sin^4\Phi_m\sin^2\Phi_{m-1}
\nonumber\\  &
+e^{i\tau(E_m^+-E_{m-1}^-)}\cos^4\Phi_m\cos^2\Phi_{m-1}+e^{i\tau(E_m^--E_{m-1}^-)}\sin^4\Phi_m\cos^2\Phi_{m-1}\big],
\end{align}
where we have also introduced the fact that $p_{m,m}=p_m^2$.\\
As we already said, according to \cite{Eberly1977}, the physical spectrum for a non-stationary processes is calculated by
\begin{equation*}
S(\nu)=\mathrm{Re}\left\{\int_{0}^\infty e^{-i\nu\tau}e^{-\gamma\tau}\bar{\Gamma}(\tau)d\tau\right \},
\end{equation*}
and after a long, but straight calculation, we get
\begin{align}\label{0320}
S(\delta)=&p_0^2\bigg[\cos^4\Phi_0\frac{\gamma}{\gamma^2+\lambda^2(\delta-c_+)^2}+\sin^4\Phi_0\frac{\gamma}{\gamma^2+\lambda^2(\delta-c_-)^2}\bigg]
\nonumber\\ &
+\sum_{m=1}^\infty p_m^2\bigg\{\cos^4\Phi_m\sin^2\Phi_{m-1}\frac{\gamma}{\gamma^2+\lambda^2[\delta-(\Lambda_m-\Lambda_{m-1})]^2}
+\cos^4\Phi_m\cos^2\Phi_{m-1}\frac{\gamma}{\gamma^2+\lambda^2[\delta-(\Lambda_m+\Lambda_{m-1})]^2}
\nonumber\\ &
+\sin^4\Phi_m\sin^2\Phi_{m-1}\frac{\gamma}{\gamma^2+\lambda^2[\delta+(\Lambda_m+\Lambda_{m-1})]^2}
+\sin^4\Phi_m\cos^2\Phi_{m-1}\frac{\gamma}{\gamma^2+\lambda^2[\delta+(\Lambda_m-\Lambda_{m-1})]^2}\bigg\}
\end{align}
with
\begin{subequations}
	\begin{align}
	\delta=&\frac{\nu-\omega}{\lambda}, \\
	c_\pm= & \frac{\Delta-\chi}{2\lambda}\pm\sqrt{\left(\frac{\Delta+\chi}{2\lambda} \right)^2+1},\\
	\Lambda_m=&\sqrt{\left[\frac{\Delta+\chi\left(2m+1\right)}{2\lambda} \right]^2+(m+1)}.
	\end{align}
\end{subequations}
According to equation \eqref{0320} the allowed transitions are
\begin{table}[H]
	\begin{center}
		\begin{tabular}{|c|c|}
			\hline
			Transition & $\delta$\\
			\hline
			$\ket{\psi_m^+}\rightarrow\ket{\psi_{m-1}^+}$ &  $\Lambda_m-\Lambda_{m-1}$\\
			\hline
			$\ket{\psi_m^+}\rightarrow\ket{\psi_{m-1}^-}$ & $\Lambda_m+\Lambda_{m-1}$\\
			\hline
			$\ket{\psi_m^-}\rightarrow\ket{\psi_{m-1}^+}$ & $-(\Lambda_m+\Lambda_{m-1})$\\
			\hline
			$\ket{\psi_m^-}\rightarrow\ket{\psi_{m-1}^-}$ & $-(\Lambda_m-\Lambda_{m-1})$\\
			\hline
			$\ket{\psi_0^+}\rightarrow\ket{0,g}$ & $c_+$\\
			\hline
			$\ket{\psi_0^-}\rightarrow\ket{0,g}$ & $c_-$\\
			\hline
		\end{tabular}
		\caption{Values of $\delta$ for which a transition occurs.}
	\end{center}
\end{table}

\section{Some examples.}
We analyze now the concrete behavior of the florescence spectrum. For that, we have to choose the parameters of the system represented by the Hamiltonian \eqref{0020} and also we need to pick an initial photon distribution for the field. In the case of the atom, we select the interaction constant of the ground $\ket{g}$ and the excited $\ket{e}$ states with the field as 1, $\lambda=1$. We have denoted the detuning between the field and the two lower states of the atom as $\Delta$, and we will take two values for such detuning, 0.0 and 0.03. The interaction constants $\eta_k$ between the ground state $\ket{g}$ and the higher nearby levels $\ket{k}$, together with the frequencies $\omega_k$, will be chosen in such a way to fix a value of the effective Stark effect parameter $\chi$; we will take for $\chi$ a zero value, which correspond to non-interaction with the higher nearby levels, and a stronger interaction, which is given by 0.9. The width of the detector will be taken always as 0.1.\\
In the case of the initial photon distribution of the field, we will pick to cases: a coherent and a thermal field. In both cases, the photon distribution is characterized by the mean photon number, $\bar{n}$, and we will examine the events when this photon mean number is equal to 1.0 and 10.0.
\subsection{Coherent field.}
\begin{figure}[H]
	\centering
	\subfloat[$\Delta=0.0, \; \chi=0.0$]
	{\includegraphics[width=0.45\textwidth]{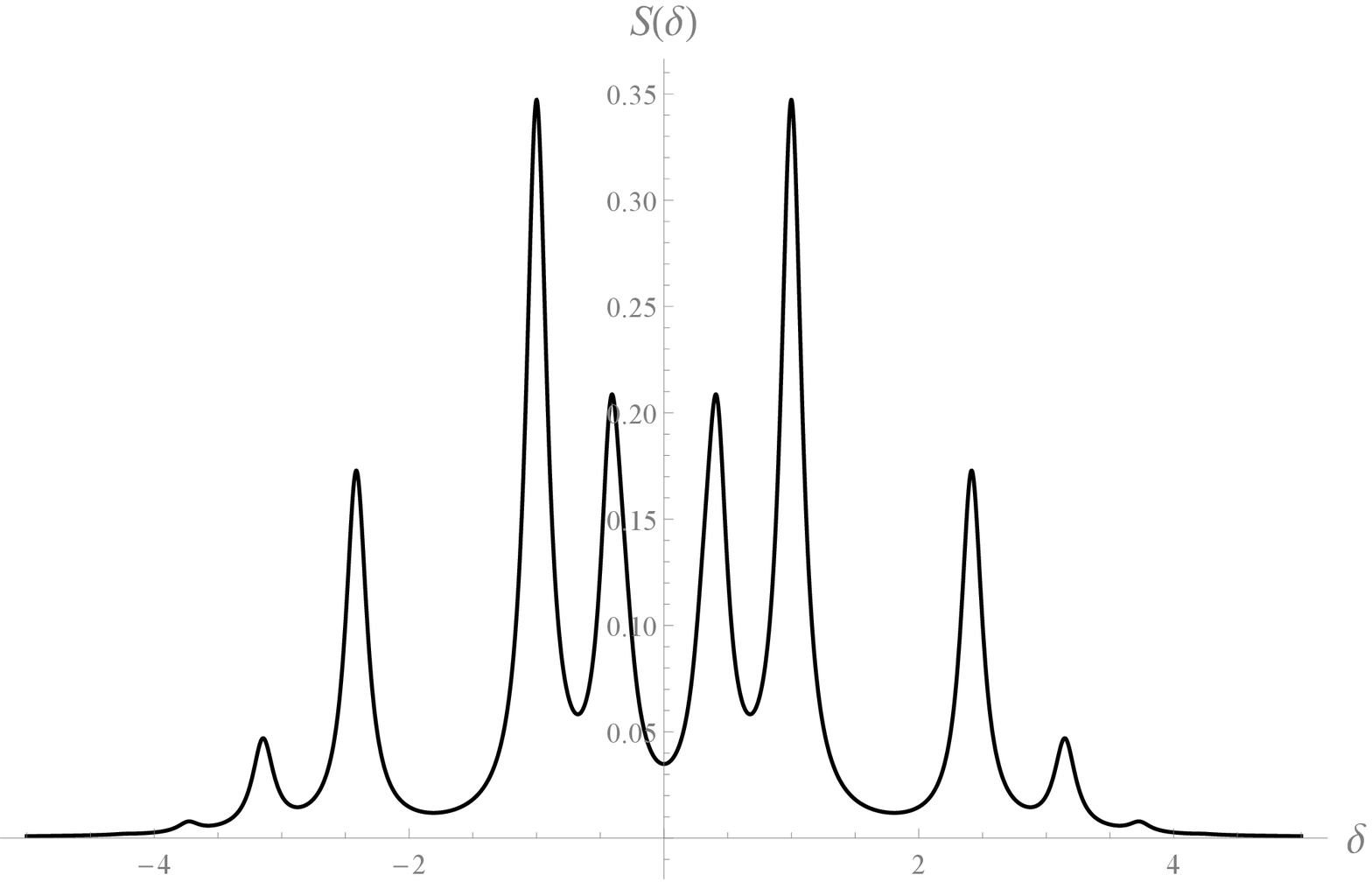}} 
	\subfloat[$\Delta=0.0, \; \chi=0.9$]
	{\includegraphics[width=0.45\textwidth]{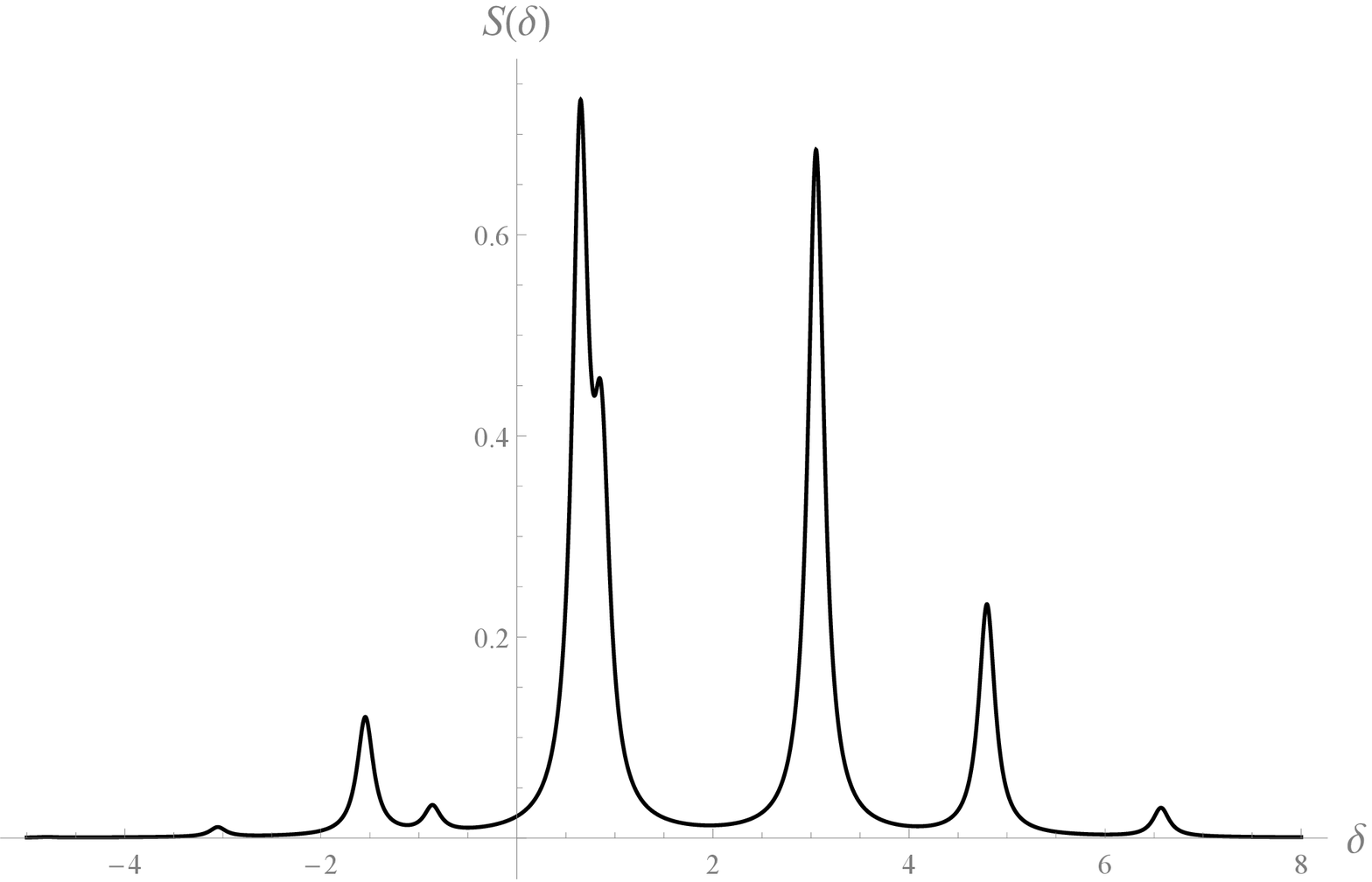}} 
	\\
	\subfloat[$\Delta=0.3, \; \chi=0.0$]
	{\includegraphics[width=0.45\textwidth]{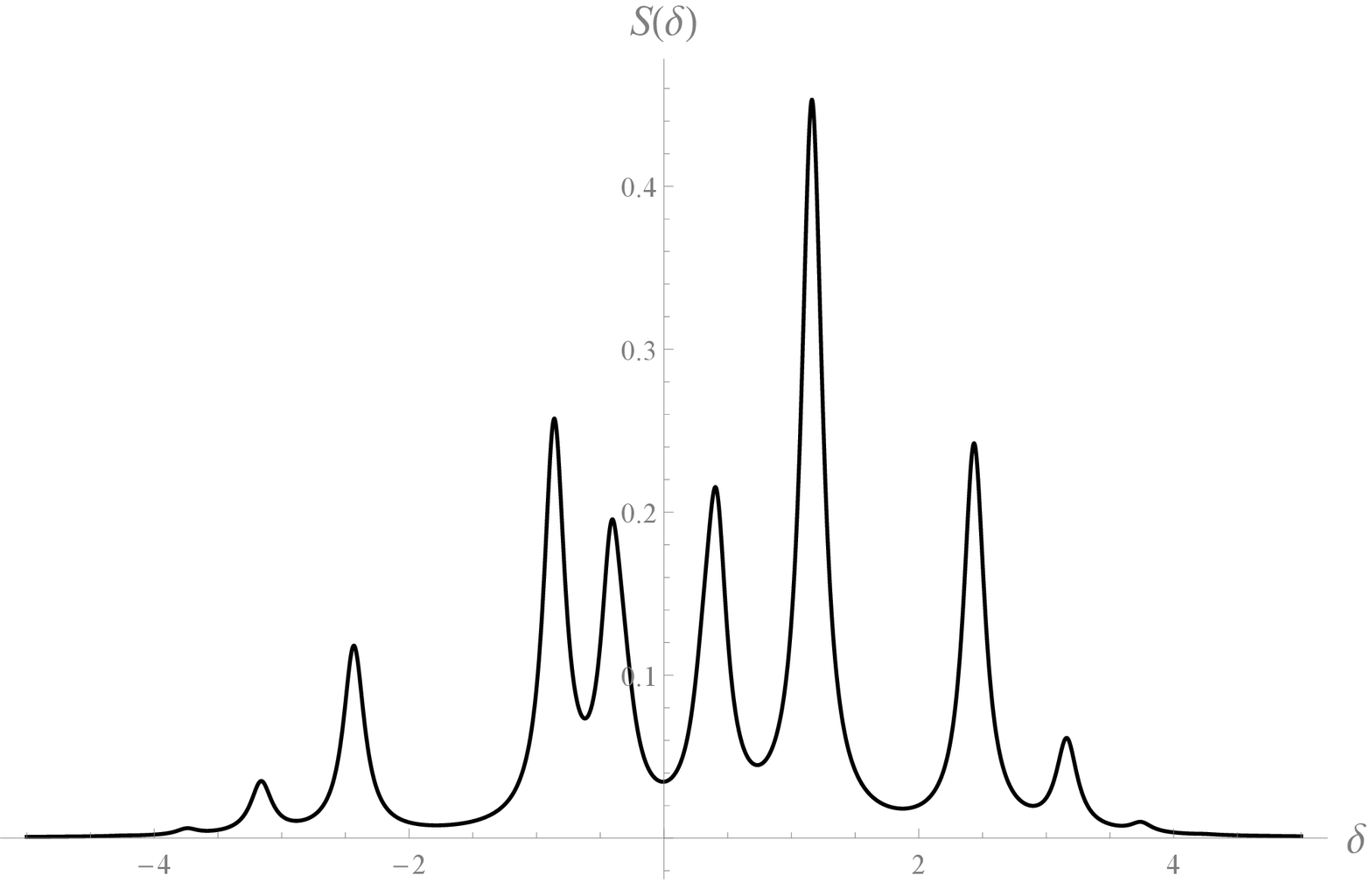}}
	\subfloat[$\Delta=0.3, \; \chi=0.9$]
	{\includegraphics[width=0.45\textwidth]{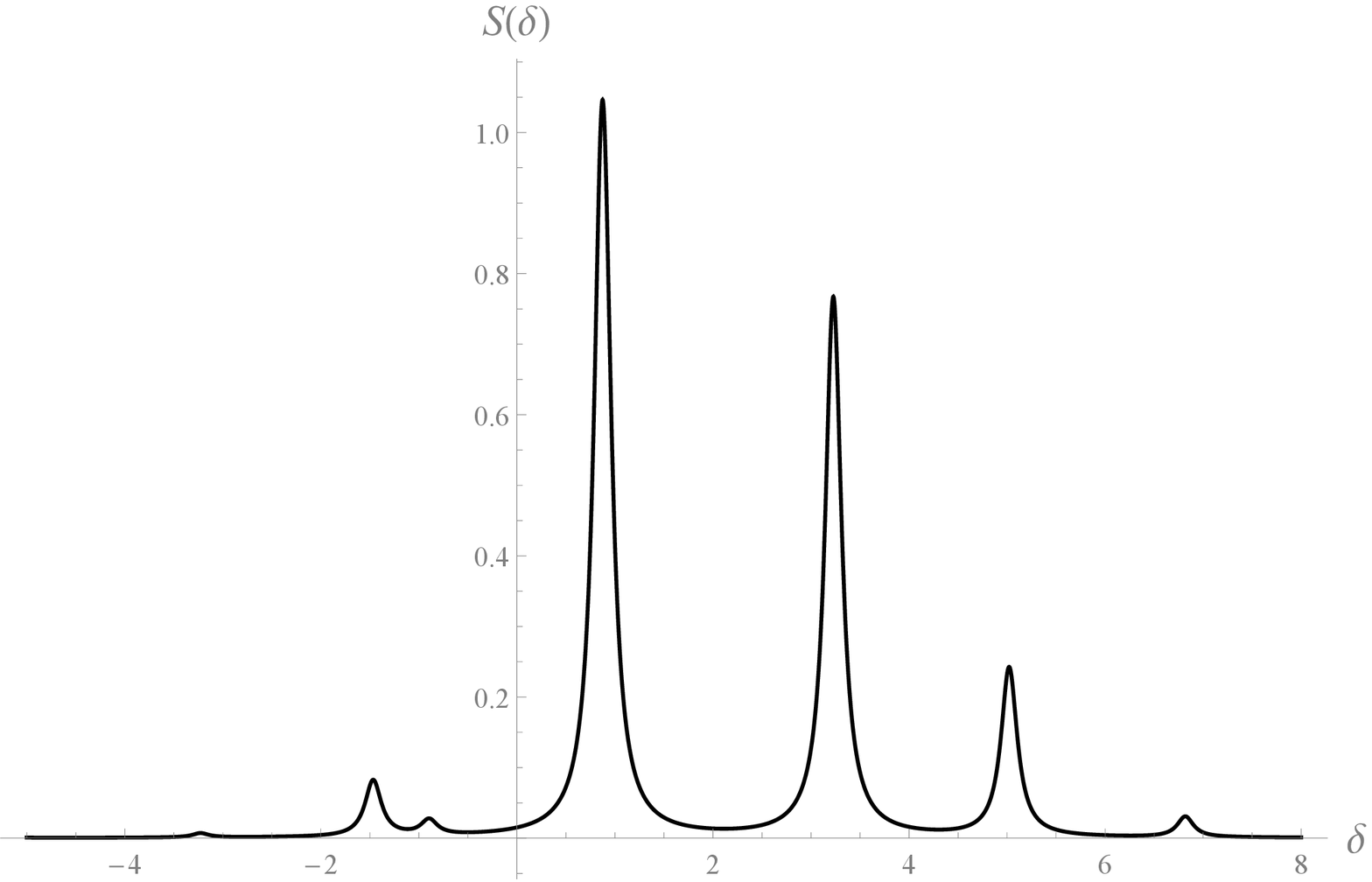}} 
	\caption{The fluorescence spectrum when the initial field is in a coherent state with $\bar{n}=1.0$}\label{fig2}
\end{figure}

\begin{figure}[H]
	\centering
	\subfloat[$\Delta=0.0, \;\chi=0.0$]
	{\includegraphics[width=0.45\textwidth]{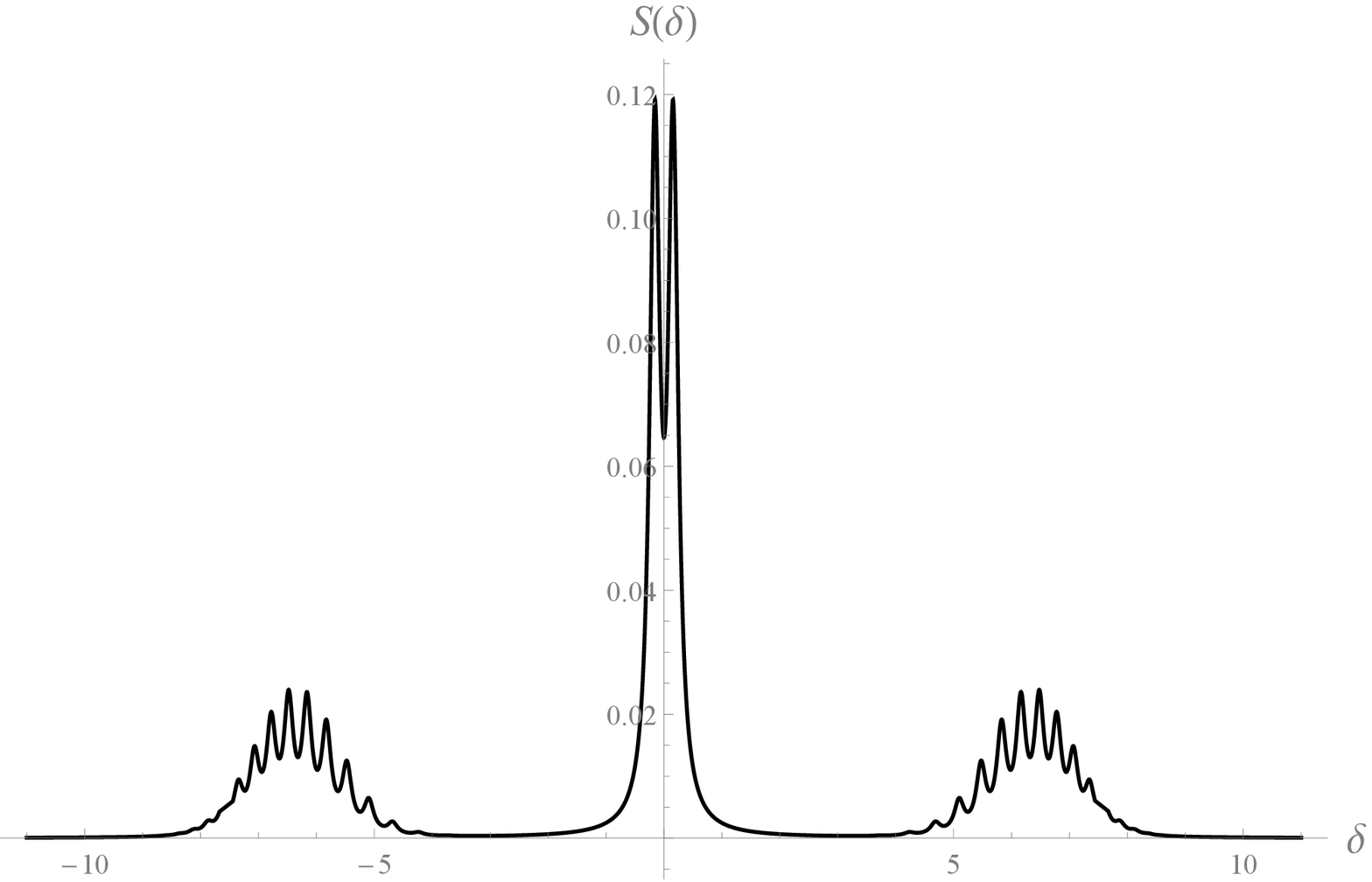}}
	\subfloat[$\Delta=0.0, \;\chi=0.9$]
	{\includegraphics[width=0.45\textwidth]{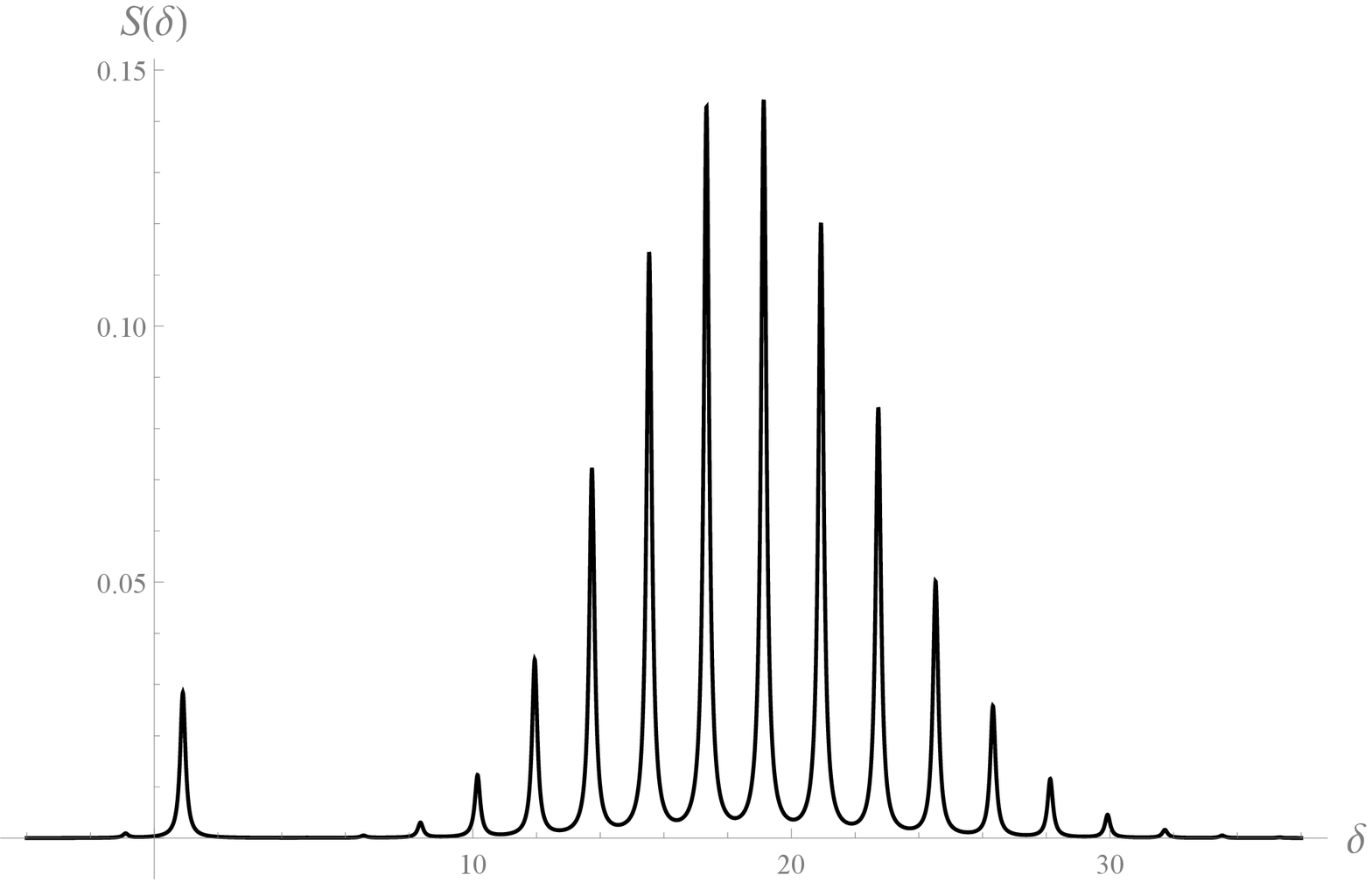}} 
	\\
	\subfloat[$\Delta=0.3, \; \chi=0.0$]
	{\includegraphics[width=0.45\textwidth]{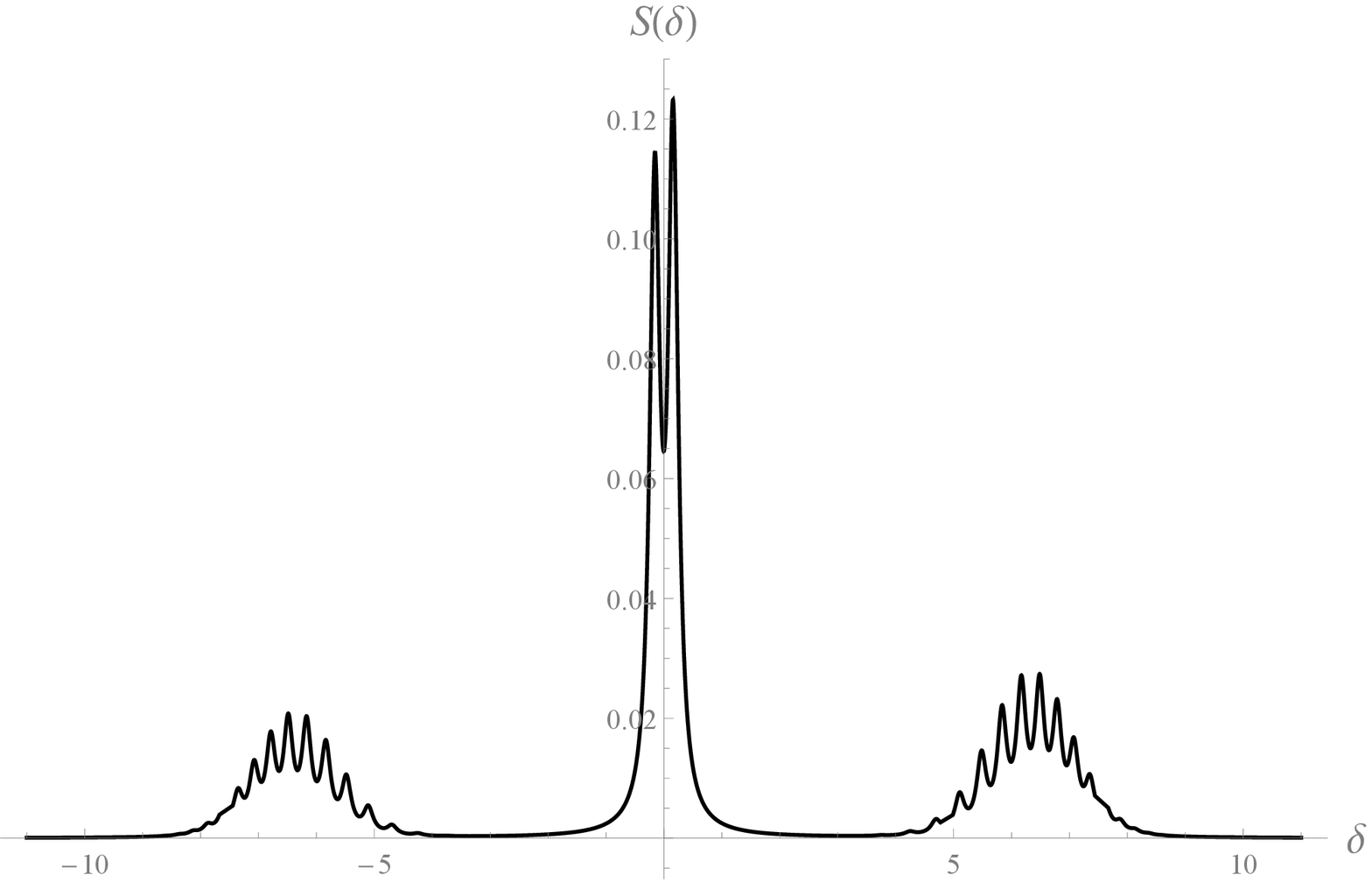}}
	\subfloat[$\Delta=0.3, \; \chi=0.9$]
	{\includegraphics[width=0.45\textwidth]{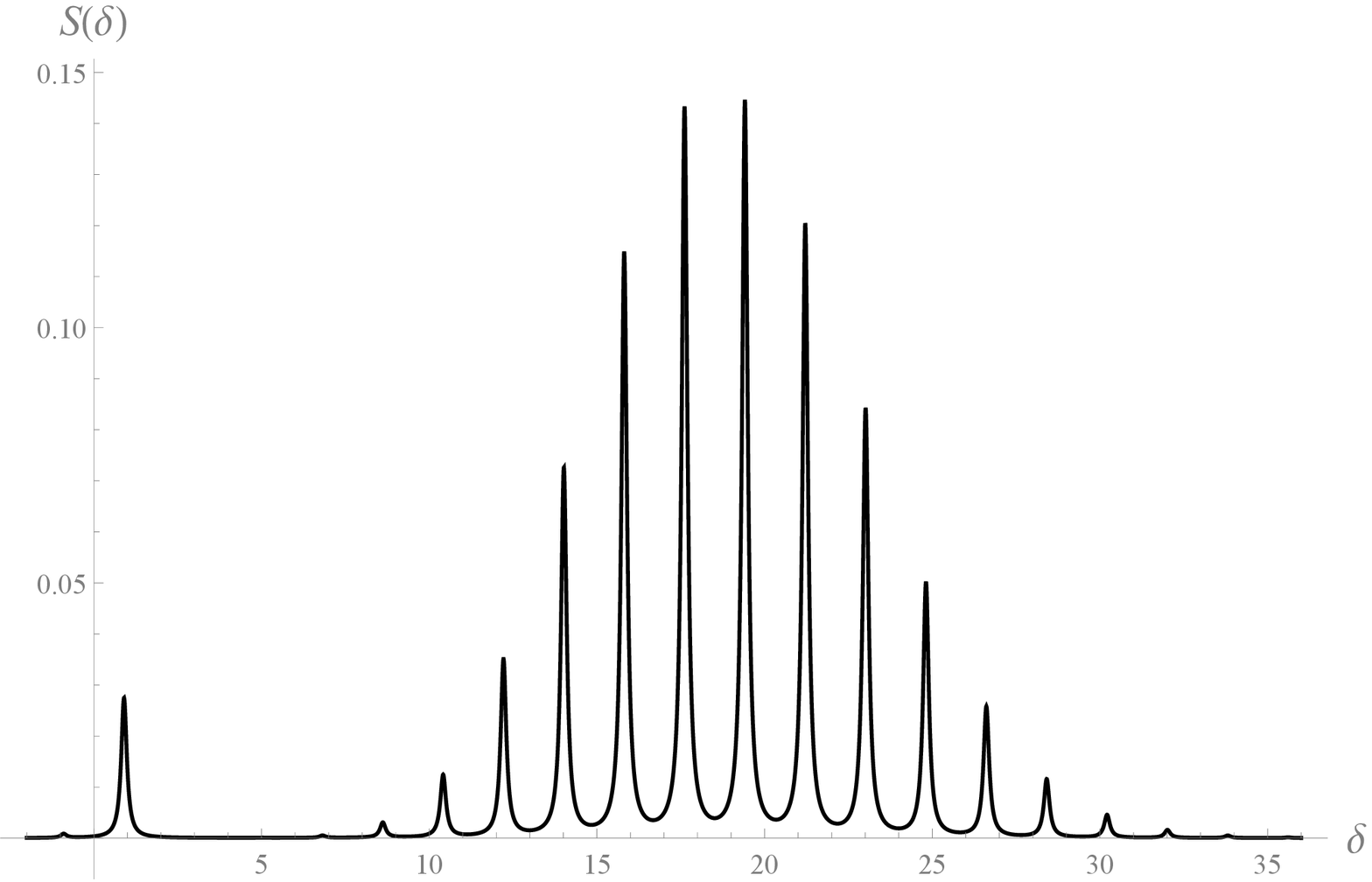}} 
	\caption{The fluorescence spectrum when the initial field is in a coherent state with $\bar{n}=10.0$}\label{fig3}
\end{figure}

The photon probability distribution of a coherent field is
\begin{equation}
p_m=\exp\left( -\bar{n} \right)\frac{\bar{n}^m}{m!}, \qquad m=0,1,2,\dots,
\end{equation}
where $ \bar{n}$ denotes the average number of photons.\\ 
For a mean photon number of 1 (always $\gamma=0.1, \; \lambda=1$), we have Fig. \ref{fig2} where it is possible to observe that when we take into account the influence of the nearby levels the spectrum becomes asymmetric and shifted with respect to the spectrum without that interaction, i.e. with $\chi=0$.
\noindent For a mean photon number of 10 (always $\gamma=0.1, \; \lambda=1$), we present the spectrum in Fig. \ref{fig3}, where the same effect that in Fig. \ref{fig2} can be observed: The interaction with the nearby levels shifts the spectrum and makes it asymmetric.

\subsection{Thermal field.}
\begin{figure}[H]
	\centering
	\subfloat[$\Delta=0.0, \; \chi=0.0$]
	{\includegraphics[width=0.45\textwidth]{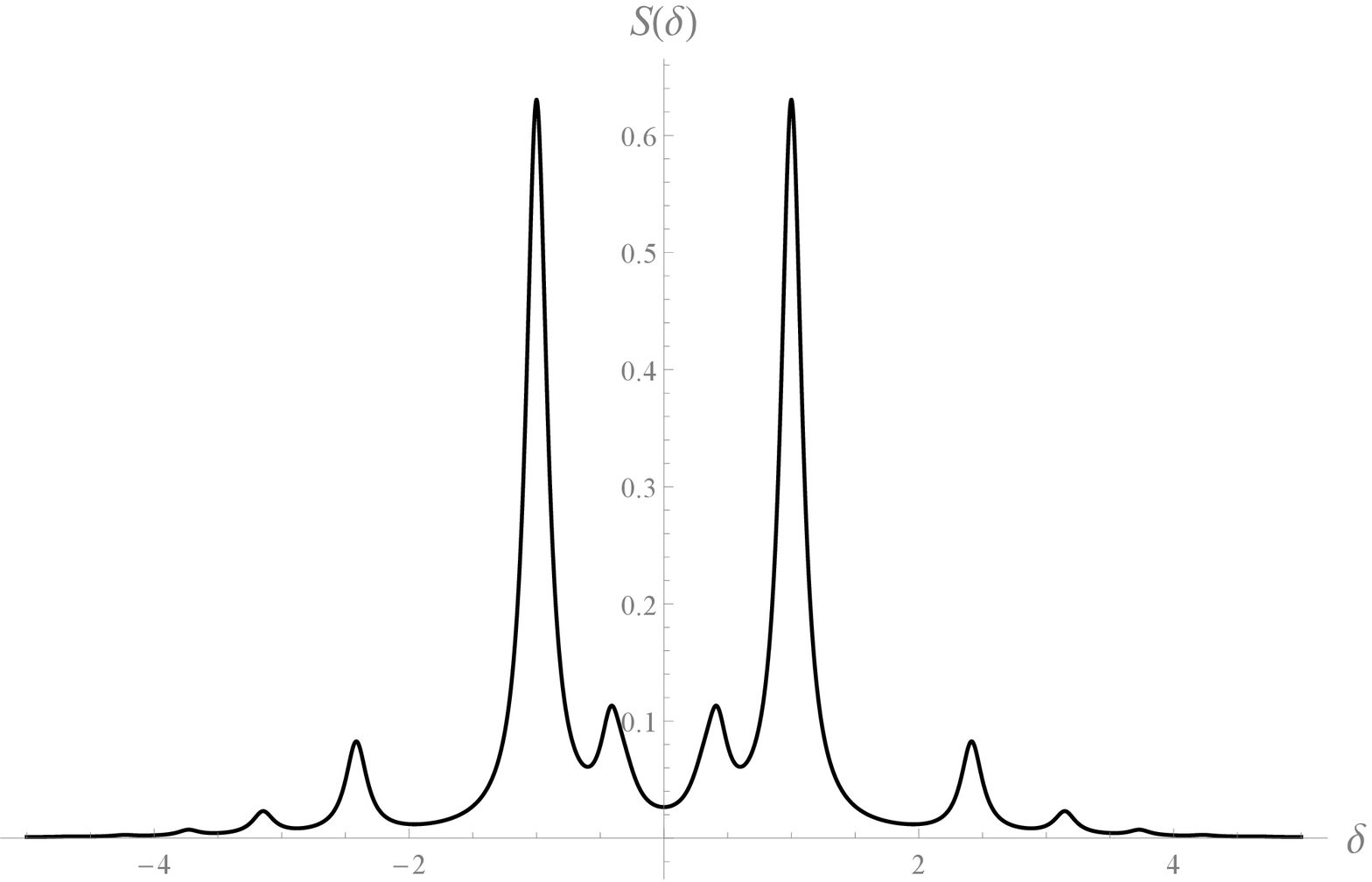}}
	\subfloat[$\Delta=0.0, \; \chi=0.9$]
	{\includegraphics[width=0.45\textwidth]{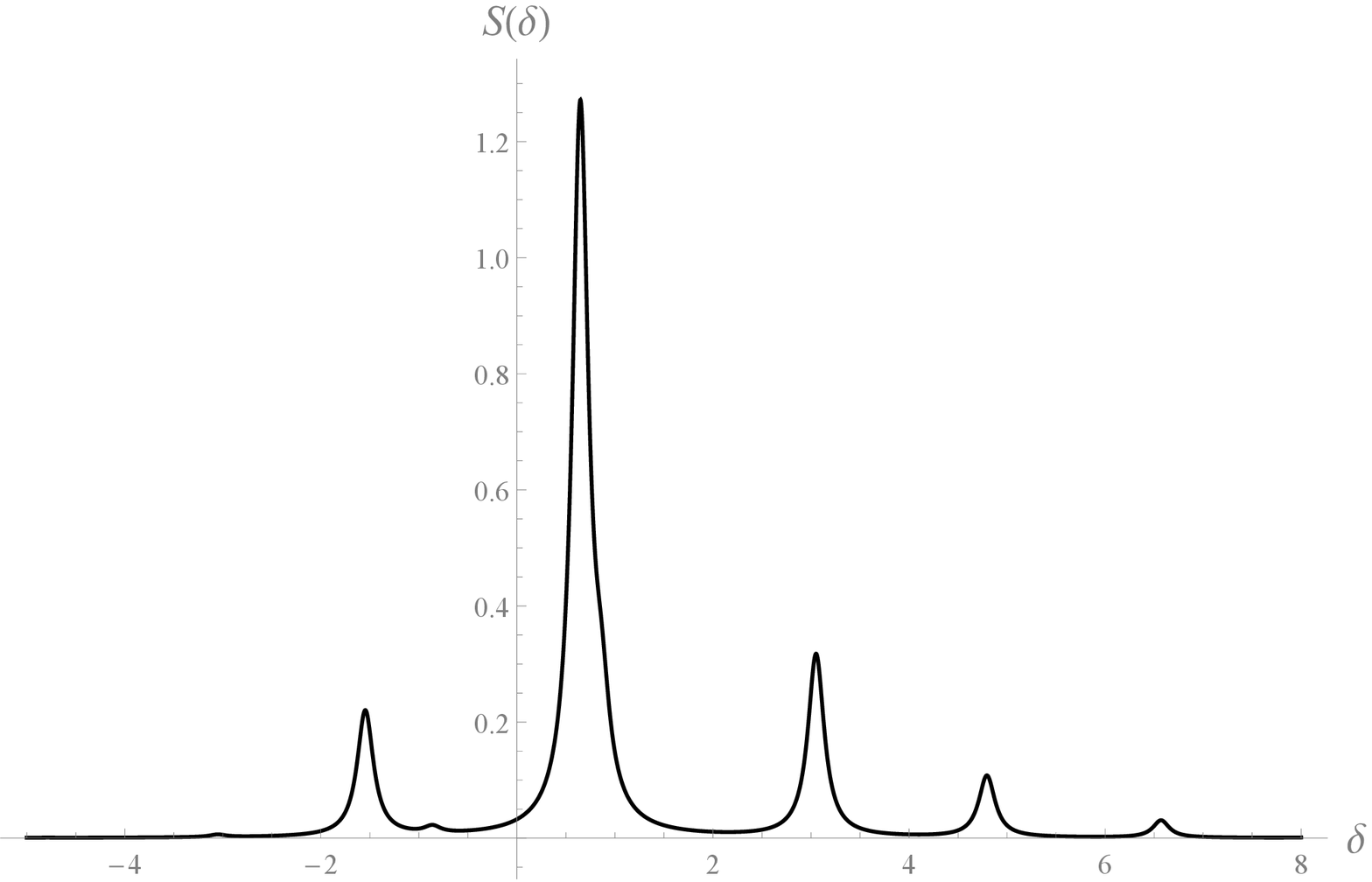}} 
	\\
	\subfloat[$\Delta=0.3, \; \chi=0.0$]
	{\includegraphics[width=0.45\textwidth]{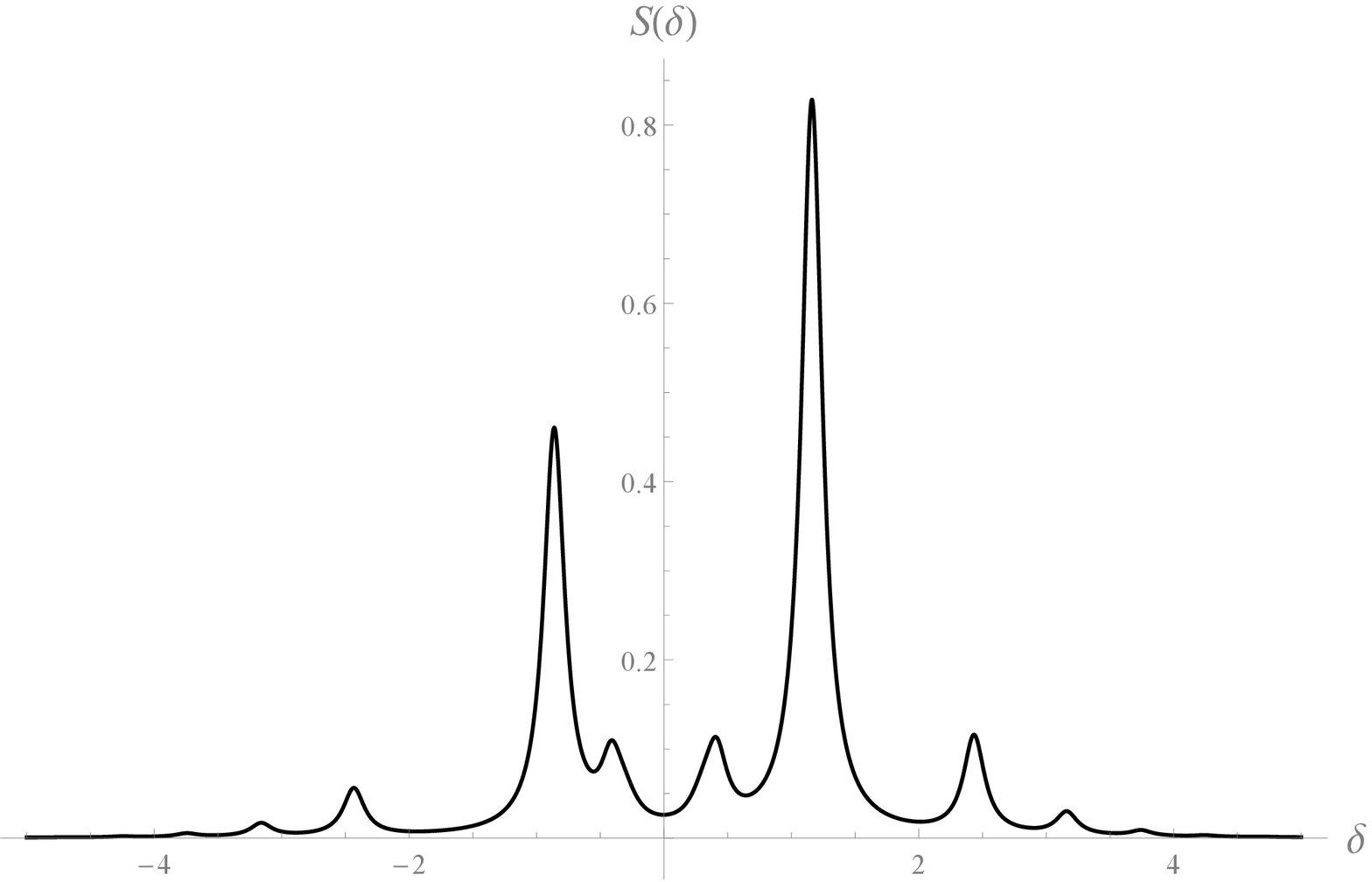}}
	\subfloat[$\Delta=0.3, \; \chi=0.9$]
	{\includegraphics[width=0.45\textwidth]{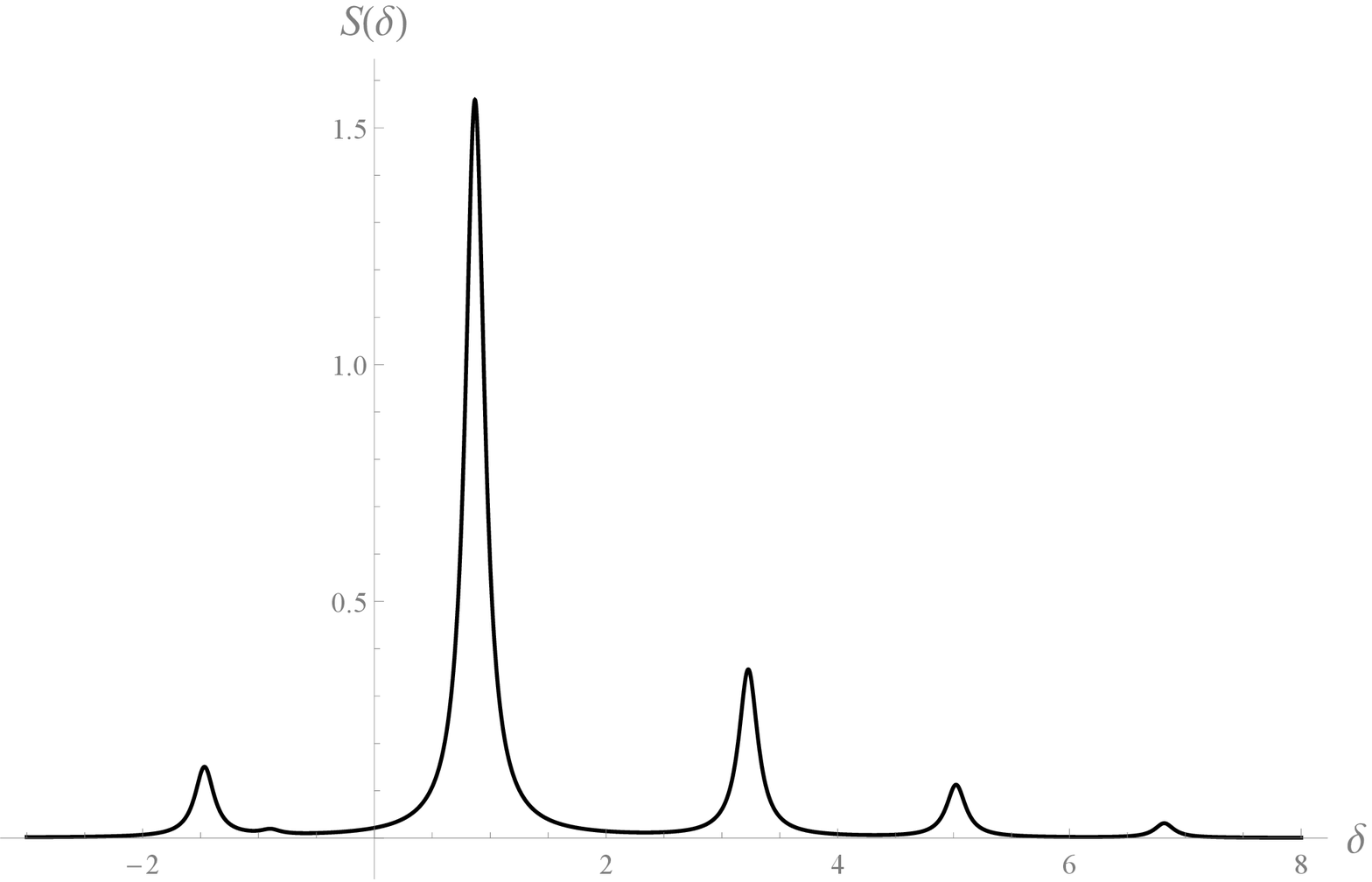}} 
	\caption{The fluorescence spectrum when the initial field is in a thermal state with $\bar{n}=1.0$}\label{fig4}
\end{figure}

\begin{figure}[H]
	\centering
	\subfloat[$\Delta=0.0, \; \chi=0.0$]
	{\includegraphics[width=0.45\textwidth]{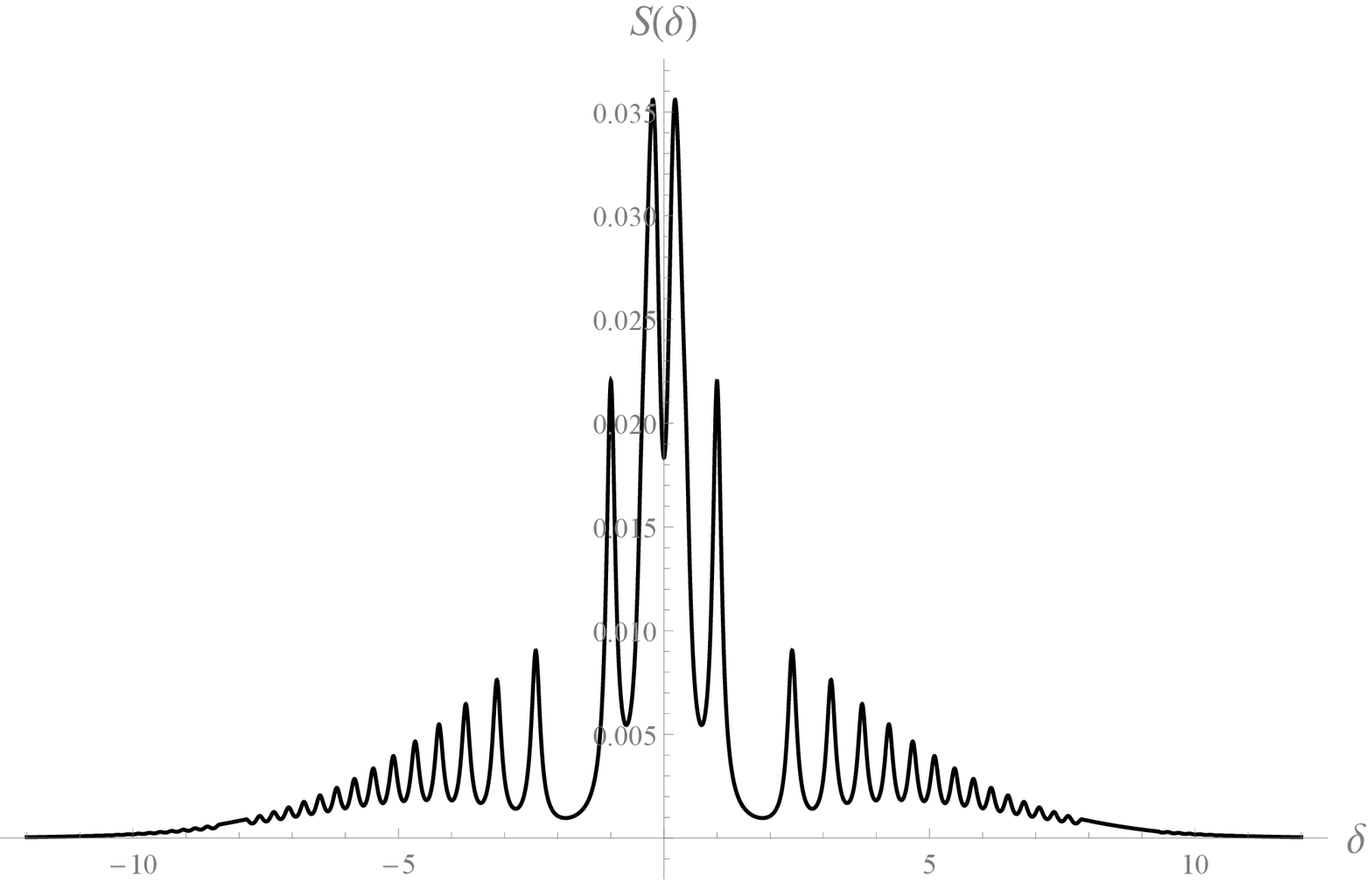}}
	\subfloat[$\Delta=0.0, \; \chi=0.9$]
	{\includegraphics[width=0.45\textwidth]{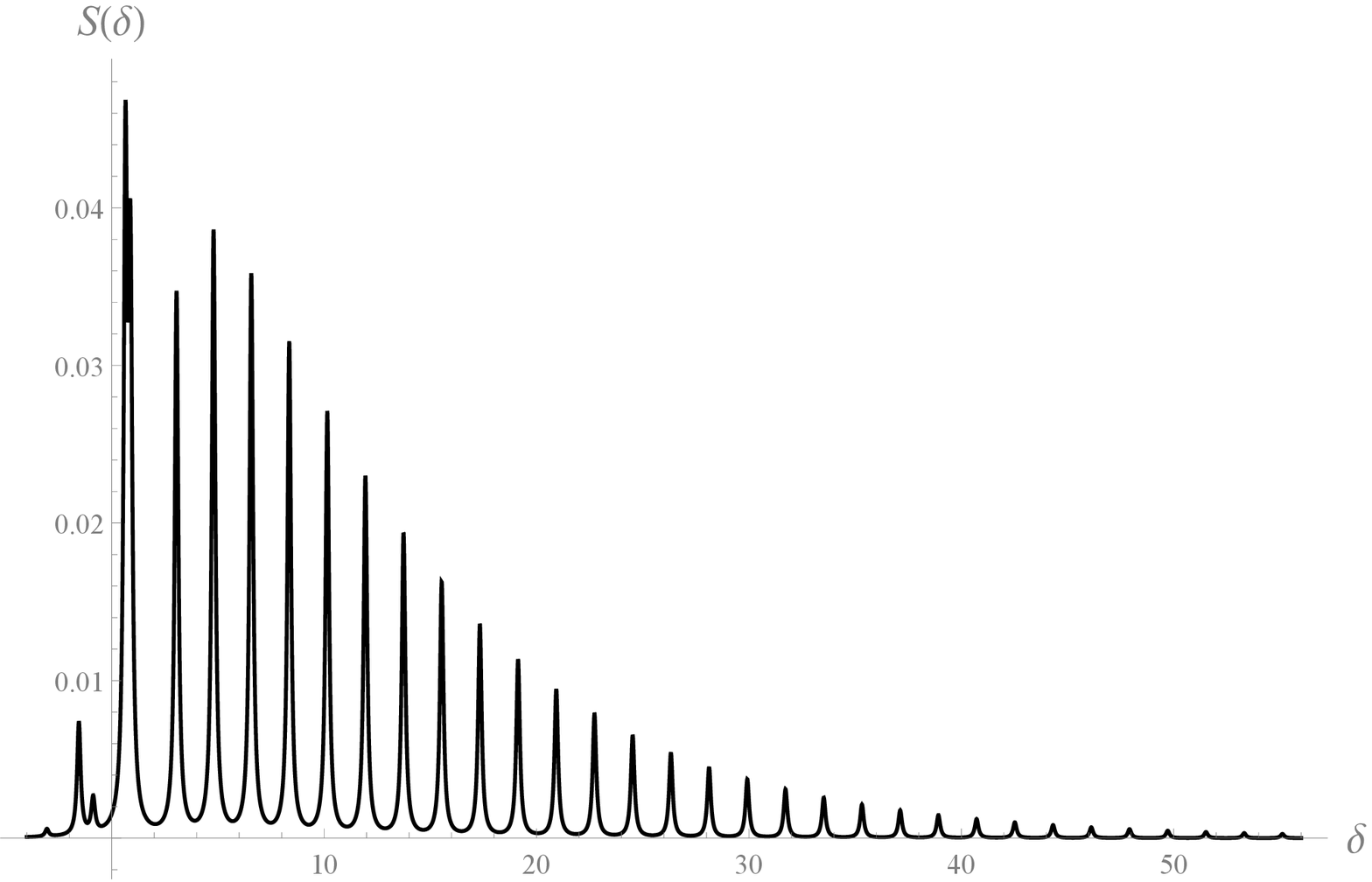}} 
	\\
	\subfloat[$\Delta=0.3, \; \chi=0.0$]
	{\includegraphics[width=0.45\textwidth]{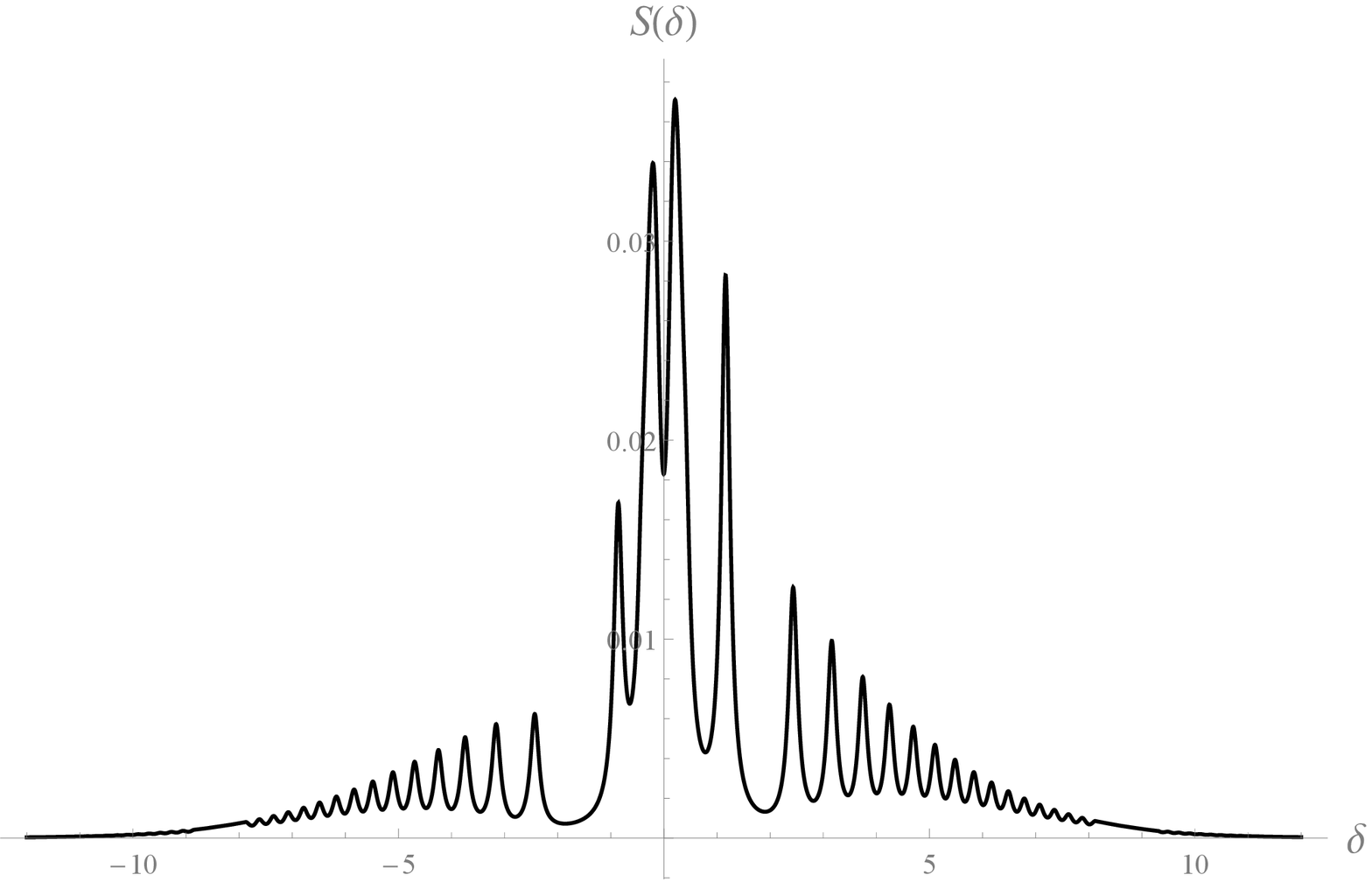}}
	\subfloat[$\Delta=0.3, \; \chi=0.9$]
	{\includegraphics[width=0.45\textwidth]{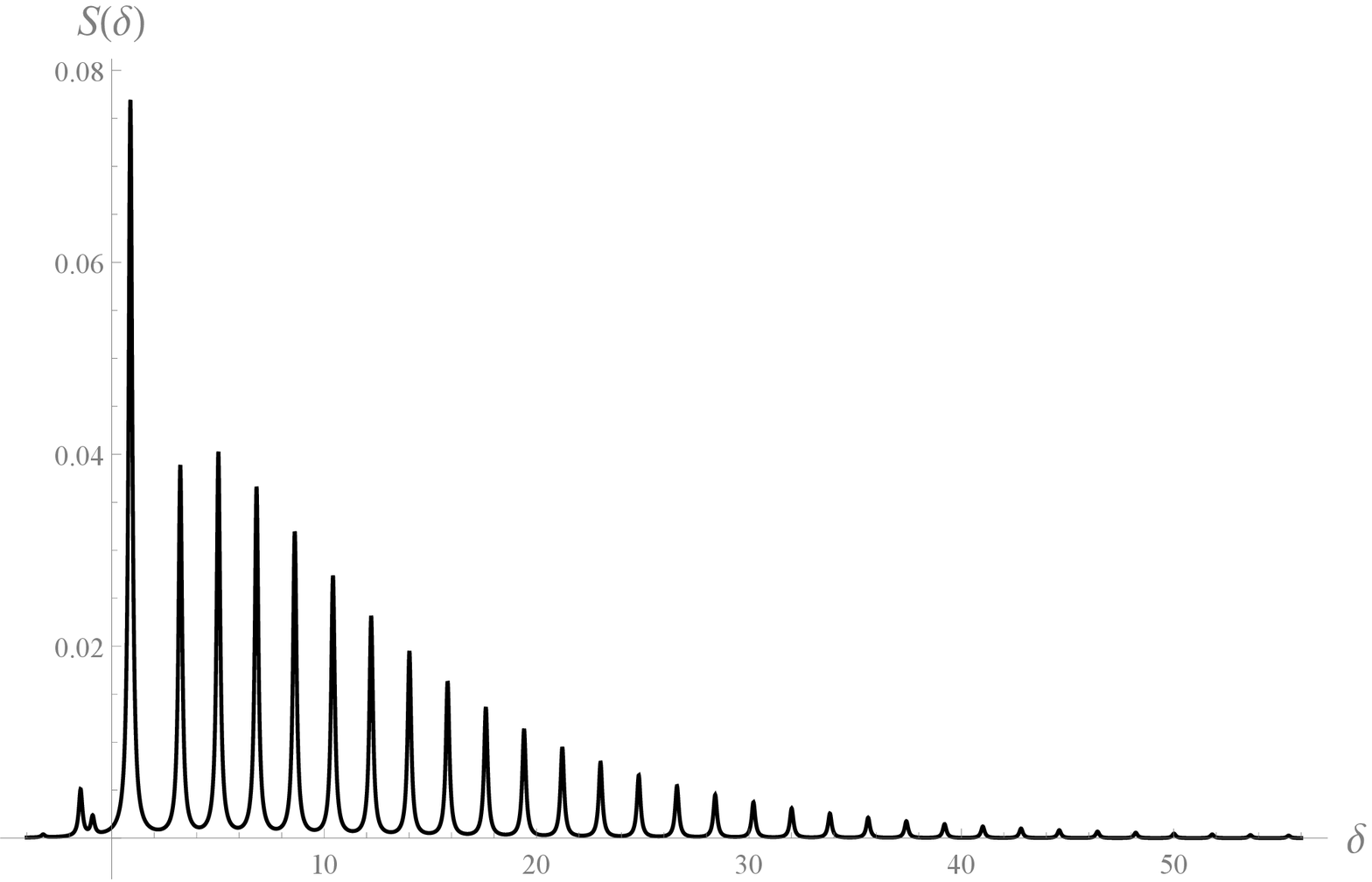}} 
	\caption{The fluorescence spectrum when the initial field is in a thermal state with $\bar{n}=10.0$}\label{fig5}
\end{figure}

Assuming we have a thermal field, the $ p_m $ function is
\begin{equation}
p_m= \frac{\bar{n}^m}{(\bar{n}+1)^{m+1}}, \qquad m=0,1,2,\dots,
\end{equation}
where $\bar{n}$ is the photon mean number.
For a mean photon number of 1 (always $\gamma=0.1, \; \lambda=1$), we observe again, in Fig. \ref{fig4}, that the non-resonant interaction with the nearby levels shifts the spectrum and makes it asymmetrical.
For a mean photon number of 10 (always $\gamma=0.1, \; \lambda=1$), we have Fig. \ref{fig5} where it appear the same effects that in the previous cases.

\section{Conclusion.} 
We show that the Hamiltonian describing the interaction between an atom of levels $ \ket{g} $, $ \ket{e} $ and higher nearby levels $ \ket{k} $ with $ k = 1 , ..., N $, can be transformed by the small rotations method into the Jaynes-Cummings Hamiltonian plus an extra term given by $ \chi\hat{n}\hat{\sigma}_z $ that depends on the number of photons inside the cavity. This term represents the effect of nearby levels on the dynamics of the two-level atom and the quantized field. According to definition \eqref{0080}, the parameter $ \chi $  contains all the information about the frequencies of the off-resonant nearby levels and an inequality related to coupling constants \eqref{0060} holds. In general, if the number of nearby levels increases, so does the value of the parameter $\chi$. Using the model described by the effective Hamiltonian $ \hat{H}_\mathrm{SE}$, we find an expression for the fluorescence spectrum or physical spectrum that consists of Lorentzian curves centered on $\delta=\Lambda_m\mp\Lambda_{m-1}, \; (\Lambda_m\pm\Lambda_{m-1})$ and $c_\pm$. In agreement with \eqref{0320}, the spectrum will be determined by the initial photon statistics of the field, detuning $ \Delta$ and parameter $\chi$. Results are presented for coherent and thermal fields with different values of $\Delta$ and $\chi$. In the event of $\Delta\neq 0$ or $\chi\neq 0$ a greater number of transitions will be carried out for $\delta >0$ for both fields. Therefore, AC Stark shifts causes asymmetries and shifts in the spectrum. In the particular,  in the case $\Delta=\chi=0 $ the spectrum is symmetric and the results reported by \cite{Agarwal1991} for the Jaynes-Cummings model are recovered.

\section{Funding}
This research did not receive any specific grant from funding agencies in the public, commercial, or
not-for-profit sectors.

\end{document}